\documentclass[12pt]{article}
\usepackage{tikz}
\usepackage{amsmath}
\usepackage{mathtools}
\usepackage{fancyhdr}
\usepackage{gensymb}
\usepackage{graphicx}
\usepackage{cite}
\usetikzlibrary{decorations.pathmorphing, arrows, patterns}
\tikzset{snake it/.style={-stealth,
decoration={snake,
    amplitude = .4mm,
    segment length = 2mm,
    post length=0.9mm},decorate}}
\usepackage{amssymb}

\usetikzlibrary{decorations.markings}

\usepackage[a4paper, total={6in, 10in}]{geometry}
\DeclareMathOperator{\ach}{Arcosh}

\begin{document}
\title{\Large{\textbf{Classical electrodynamics on Snyder space}}}
\date{}
\author{Boris Iveti\'c
\footnote{bivetic@yahoo.com}
\\University of Vienna, Faculty of Physics
  \\Boltzmanngasse 5, 1090 Vienna, Austria
 \\ Department of Physics, Faculty of Science, University of Zagreb
  \\ Bijeni\v cka cesta 32, 10000 Zagreb, Croatia
}

\maketitle

\begin{abstract}
A formulation of classical electrodynamics on an energy-momentum background of constant, non-zero curvature is given. The procedure consists of taking the formulation of standard electrodynamics in the energy-momentum representation, and promoting the energy-momentum vector to belong to a constant (non-zero) curvature space. In particular, special emphasis is given to the definition of integration measure and generalized Dirac's delta function. Finally, simple physical problems as plane waves (solutions outside sources) and point charges are discussed in this context, where the  self-energy of a point charge is shown to be finite.
\end{abstract}

\section{Introduction}
Field theory on the background of an  energy-momentum space with constant curvature has been studied extensively in the quantum case \cite{golf1, Golf, mk1, mk2, kady1, kady2, kady3, kady4, mk3, mk4}. While QFT is quite developed, classical field theory has not been studied specifically. The deviation from the zero-curvature energy-momentum background is a generic proposal that should be applicable to both classical and quantum cases, as well as to fields and point-particles (for the latter, see recent efforts in e.g. \cite{relloc,mign1}). Here it is applied to classical electrodynamics, leading to a formulation that preserves both Lorentz and gauge symmetries, since the Lorentz and gauge groups have standard representations on the background of constant curvature energy-momentum space (as first noted in case of Lorentz symmetry by Snyder \cite{Snyder1}). As in the quantum case, where it led to finite amplitudes at every step \cite{mk4}, in classical case the nontrivial integration measure proves again to be a crucial property, leading to a finite self-energy of point charge.

Historically, the first emergence of nontrivial geometry of the energy-momentum space came in the context of noncommutativity of space-time, in the paper by Snyder in 1947 \cite{Snyder1}. This was related to the existence of a minimal uncertainty of the position, i.e. a minimal length, which was to be the cure for the divergences in the developing field theory. Due to the success of renormalization methods the problem did not attract much attention, until its reemergence in the 1990-ies with the development of string theory and quantum gravity models, which led to the demand of modification of the structure of space-time at the smallest scales, see e.g. \cite{st,qg}. At the same time, the theory of quantum groups and quantum (noncommutative) geometry saw rapid development as mathematical disciplines \cite{books}. In addition to Snyder's model of noncummutativity, there appeared also other models of noncommutativity, most notably the kappa-Minkowski model \cite{kappa} and the canonical model \cite{can}, as well as theories with modified relativity, such as DSR model \cite{dsr}, which all have in common the nontrivial geometry of the corresponding energy-momentum spaces. In this paper, we shall focus on Snyder's model and the resulting de Sitter geometry of the momentum space, since de Sitter space is geometrically the simplest generalization of the flat space, and unlike in the mentioned kappa-Minkowski and canonical noncommutativity, and DSR model, the fundamental Lorentz symmetry is not deformed.

The plan of the paper is as follows. In Section 2 the main principles of the formulation of standard classical electrodynamics  on energy-momentum background are reviewed, and the principle of minimal extension to the energy-momentum space of non-zero curvature is introduced. Section 3 studies some elements of the geometry of constant curvature space, and generalizes the concept of Dirac's delta function on such spaces. Section 3 proceeds directly to the application of curvature on the energy-momentum space in physical problems of plane waves and point charges. Section 4 summarizes the paper and gives outlook for future research.
\section{Classical electrodynamics in energy-momentum representation}
The usual way to study electrodynamics is on  space-time. But it can also be formulated on the energy-momentum space. For this purpose the action principle is used, which is shortly reviewed below.

\subsection*{Pure gauge (source-free)}
In the absence of sources, the familiar action on space-time is\footnote{In general, we write vectors and tensors without indices. The scalar product of vectors is denoted by a parenthesis, $(xp)\equiv x^\mu p_\mu$,\ \ $x^2=x^\mu x_\mu$, and similarly for tensors, $(FG)=F^{\mu\nu}G_{\mu\nu}$,\ \ $F^2=F^{\mu\nu}F_{\mu\nu}$.}

\begin{equation}
S_0=-\frac{1}{4}\int F^2(x) d^dx,
\end{equation}
where $F(x)$ is the electromagnetic field tensor, with components

\begin{equation}
F_{\mu\nu}(x)\equiv \partial_\mu A_\nu(x)-\partial_\nu A_\mu(x),
\end{equation}
and $A(x)$ is a vector potential in $d$ dimensions. The form of the action is dictated by demands of Lorentz and gauge invariance, which are independent of the representation. Therefore on the energy-momentum space the action is of the same form \cite{kady4}

\begin{equation}
S_0=\pi\int \left(F^\dagger(p) F(p)\right) d^dp,
\end{equation}
where $F(p)$ is the electromagnetic field tensor in the energy-momentum representation, with components

\begin{equation}\label{field tensor emom}
F_{\mu\nu}(p)\equiv p_\mu A_\nu(p)-p_\nu A_\mu(p),
\end{equation}
and $A(p)$ is Fourier transform of the vector potential,

\begin{equation}\label{fourier}
A(p)=\frac{1}{(2\pi)^{d/2}}\int A(x) e^{ipx}d^dx,
\end{equation}
with the property $A^\dagger(p)=A(-p)$

The Lagrangian density is a complete square\footnote{Technically it is necessary to preform Wick's rotation first.}. Putting it in the form

\begin{equation}\label{lag den mom}
S_0=\pi\int \frac{d^dp}{p^2}(p^2A-p (pA))^\dagger(p^2A-p (pA)),
\end{equation}
the minimum of the action immediately returns Maxwell's equations,

\begin{equation}\label{momentum maxwell standard}
S_{min}=0\Rightarrow p^2A(p)-p(pA(p))=0.
\end{equation}
Upon inverse Fourier transformation of (\ref{momentum maxwell standard}), the usual expressions for space-time description are recovered

\begin{equation}\label{spacetime maxwell standard}
\partial^2A(x)-\partial(\partial A(x))=0,
\end{equation}
or in terms of the field strength,

\begin{equation}
\partial^\nu F_{\mu\nu}(x)=0.
\end{equation}

In passing to an energy-momentum background of constant non-zero curvature, the electromagnetic field tensor in energy-momentum representation is taken to remain as in (\ref{field tensor emom}), with the understanding that all the vectors belong to a constant curvature space. The action is the same as in (\ref{lag den mom}), with only integration measure changed, to be discussed in the next section. By the same reasoning as above, equation (\ref{momentum maxwell standard}) is derived for the constant curvature energy-momentum space.

These considerations form the basis of what we call the \textit{minimal extension principle}: in passing to electrodynamics on energy-momentum background of constant (non-zero) curvature, all the expressions from the flat case are taken, with energy-momenta promoted to vectors on a surface of constant curvature. Below is explained how this principle works for gauge symmetry, and in some simple physical problems.

%
%

\subsection*{With sources}
In this case the action can be written as

\begin{equation}
S_0=\pi\int \frac{d^dp}{p^2}(p^2A-p (pA)-j)(p^2A-p (pA))^\dagger,
\end{equation}
where $j=j(p)$ is a $d$-dimensional current vector in the momentum representation, and current conservation is expressed as $(pj)=0$.

This leads to

\begin{equation}
S_{min}=0\Rightarrow p^2A(p)-p(pA(p))=j(p),
\end{equation}
and upon Fourier transformation, in coordinate representation,

\begin{equation}\label{spacetime maxwell standard s}
\partial^2A(x)-\partial(\partial A(x))=j(x),
\end{equation}
or
\begin{equation}
\partial^\nu F_{\mu\nu}=j_\mu(x)
\end{equation}

\subsection*{Gauge symmetry}
The invariance under gauge transformations is usually studied on space-time. It manifests itself in the invariance of equations (\ref{spacetime maxwell standard}) and (\ref{spacetime maxwell standard s}) for the addition  to the vector potential of a derivative of an arbitrary scalar function, i.e.

\begin{equation}
A(x)\to A(x)+\partial \Lambda(x).
\end{equation}
The action of the gauge group can be also observed on the energy-momentum space. Here it manifests itself in the invariance of the equation (\ref{momentum maxwell standard}) under the addition of an arbitrary vector function to the potential,

\begin{equation}\label{gauge momentum standard}
A(p)\to A(p)+ip\Lambda(p),
\end{equation}
where

\begin{equation}
\Lambda(p)=\frac{1}{(2\pi)^{d/2}}\int \Lambda(x) e^{-ipx}d^dx
\end{equation}
and $\Lambda^\dagger(p)=\Lambda(-p)$.
It is obvious that the transformation (\ref{gauge momentum standard}) leaves eq.~(\ref{momentum maxwell standard}) invariant regardless of the geometry of energy-momentum manifold. This means that the action of the gauge group on the energy momentum manifold is the standard one even if the energy-momentum space has a non-vanishing constant curvature.

\section{Energy-momentum space of constant curvature}
Here some elements of the geometry of energy-momentum space of constant curvature, that were studied in \cite{ourpaper}, are briefly reviewed, and some results improved to all order in the curvature parameter. A $d$-dimensional space of constant curvature is realized as a surface embedded in $d+1$ dimensional Euclidean background, specifically a hyperboloid,

\begin{equation}
\eta_0^2-\eta_1^2-\cdots-\eta_{d-1}^2-\eta_d^2=-\beta^{-2}
\end{equation}
where $\beta$ is a constant of dimension of length. The physical origin of this constant (its relation with other fundamental constants) has been considered in \cite{relloc, markov}, but at this point it can be taken as an independent constant describing the curvature of the energy-momentum manifold. Physical degrees of freedom are conveniently described by projective coordinates. The most general symmetry preserving projection is defined by

\begin{equation}
p=g(\beta^2\eta^2)\eta,
\end{equation}
with $g$ an arbitrary function and $\eta=(\eta_0,\eta_1,\dots,\eta_{d-1})$. The inverse is given by

\begin{equation}
\eta=h(\beta^2p^2)p, \ \ \ \ \eta_d=\beta^{-1} \sqrt{1+\beta^2p^2h^2},
\end{equation}
with $gh=1$.

The line element follows from the equation of the surface in embedding coordinates,

\begin{equation}
ds^2=d\eta^2-d\eta_d^2.
\end{equation}
From the defining relations it follows

\begin{equation}
d\eta=\left(\frac{\partial\eta}{\partial p}dp\right)=hdp+2\beta^2h'(pdp)p,
\end{equation}
where $h'=\partial h/\partial (\beta^2p^2)$,  as well as

\begin{equation}
d\eta_d=\frac{\beta h(h+2\beta^2p^2h')}{\sqrt{1+\beta^2p^2h^2}}(pdp),
\end{equation}
which combine into

\begin{equation}\label{general projection infinitesimal metric}
ds^2=h^2dp^2+\beta^2 \frac{4h'(h+\beta^2p^2h')-h^4}{1+\beta^2p^2h^2} (pdp)^2,
\end{equation}
where $h'=\partial h/ \partial(\beta^2p^2)$. Hence the metric tensor is

\begin{equation}
g(p)=h^2\mathbf 1_-+\beta^2 \frac{4h'(h+\beta^2p^2h')-h^4}{1+\beta^2p^2h^2}p\otimes p,
\end{equation}
with notation $\mathbf 1_-=\text{diag}(-1,1,\dots,1)$, and the symbol $\otimes$ denoting tensor or outer product, that is for the two vectors $p=(p_0,p_1,\dots ,p_{d-1})$ and $q=(q_0,q_1,\dots, q_{d-1})$

\begin{equation}
p\otimes q=
\begin{pmatrix}
  p_0q_0 & p_0q_1 & \cdots & p_0q_{d-1} \\
  p_1q_0 & p_1q_1 & \cdots & p_1q_{d-1}\\
  \vdots & \vdots & & \vdots\\
  p_{d-1}q_0 & p_{d-1}q_1 & \cdots & p_{d-1}q_{d-1}
 \end{pmatrix}.
\end{equation}
An infinitesimal volume element (in $d$ dimensions) is written as

\begin{equation}
d\Omega_p=\sqrt{|\text{det}g|}d^dp=h^{d-1} \frac{h+2h'\beta^2p^2}{\sqrt{1+\beta^2p^2h^2}}d^dp.
\end{equation}

Besides the metric tensor, the geometry of a certain space can be described with the distance function. For the two points $\eta$ and $\nu$ on the hyperboloid, the distance function is the geodesic distance between them

\begin{equation}\label{distance f embedding hyp}
d(\eta, \nu)=\beta^{-1}\ach(\beta^2(\eta_0\nu_0-\eta_1\nu_1-\dots-\eta_d\nu_d)).
\end{equation} 
On the projected space, where the points $\eta$ and $\nu$ project to points $p$ and $q$ respectively, the distance function is

\begin{equation}
d(p,k)=\beta^{-1}\ach \left(h_p h_k pk - \beta^2 \sqrt{1+\beta^2 h_p^2p^2}\sqrt{1+\beta^2 h_k^2 k^2}   \right),
\end{equation}
 where $h_p$ and $h_k$ are $h(\beta^2 p^2)$ and $h(\beta^2 k^2$). 

Infinitesimal generators of the group of isometries of de-Sitter space form two subgroups: rotations, with infinitesimal generators with components

\begin{equation}
\hat J_{\mu\nu}=\eta_\mu\frac{\partial}{\partial \eta_\nu}-\eta_\nu\frac{\partial}{\partial \eta_\mu}=p_\mu\frac{\partial}{\partial p_\nu}-p_\nu\frac{\partial}{\partial p_\mu},
\end{equation}
and displacements, with infinitesimal generators

\begin{align}
\hat x & =\beta\left( \eta_4\frac{\partial}{\partial \eta}-\eta\frac{\partial}{\partial\eta_4}\right)=\beta\eta_4  \frac{\partial}{\partial \eta}\\
&= \frac{\sqrt{1+h^2\beta^2p^2}}{ h} \left[ \frac{\partial}{\partial p}- \frac{2\beta^2h'}{h+2\beta^2p^2h'}p\left(p\frac{\partial}{\partial p}  \right)        \right]\\
&\equiv f_1\frac{\partial}{\partial p}+f_2\beta^2p\left(p\frac{\partial}{\partial p}  \right),
\end{align}
where in the last line functions $f_1(\beta^2p^2)$ and $f_2(\beta^2p^2)$, satisfying

\begin{align}
f_1&=\frac{\sqrt{1+h^2\beta^2p^2}}{ h}, \ \ \  h=\frac{1}{\sqrt{f_1^2-\beta^2p^2}} \\
f_2&=\frac{-2h'\sqrt{1+h^2\beta^2p^2}}{ h(h+2\beta^2p^2h')} =   \frac{2f_1f_1'-1}{f_1-2\beta^2p^2f_1'},
\end{align}
are introduced. The group of isometries $O(1,n)$ defines a Lie algebra through the deformed Poisson brackets between displacements. A full Poisson algebra between components of (global) displacements and momenta is 

\begin{equation}
\lbrace\hat x_\mu,  p_\nu\rbrace=f_1\eta_{\mu\nu}+\beta^2f_2p_\mu p_\nu
, \ \ \ \lbrace\hat x_\mu, \hat x_\nu\rbrace=\beta^2\hat J_{\mu\nu},\ \ \ \lbrace p_\mu, p_\nu \rbrace =0\ ,
\end{equation}
while the local algebra is undeformed,

\begin{equation}
\lbrace x_\mu,  p_\nu\rbrace=\eta_{\mu\nu}
, \ \ \ \lbrace x_\mu,  x_\nu\rbrace= \lbrace p_\mu, p_\nu \rbrace =0\ ,
\end{equation} 
where we denote by $x=x(p)=\partial/\partial p$ (see eq. (\ref{x})) the (local) canonical coordinates, the momenta of the momenta.

Finite displacements (and rotations) are generated by successive application of infinitesimal ones. For the case of homogeneous coordinates, first considered by Snyder \cite{Snyder1}

\begin{equation}\label{sphere hom proj}
p=\frac{\eta}{\beta\eta_d}=\frac{\eta}{\sqrt{1+\beta^2\eta^2}}
\end{equation}
with inverse relations

\begin{equation}\label{sphere hom proj inv}
\eta=\frac{p}{\sqrt{1-\beta^2p^2}}, \ \ \ \eta_d=\frac{1}{\beta\sqrt{1-\beta^2p^2}},
\end{equation}
finite displacements of a point $p$ by a point $k$ are  (see fig 1)

\begin{figure}
\begin{center}
\includegraphics[scale=0.5]{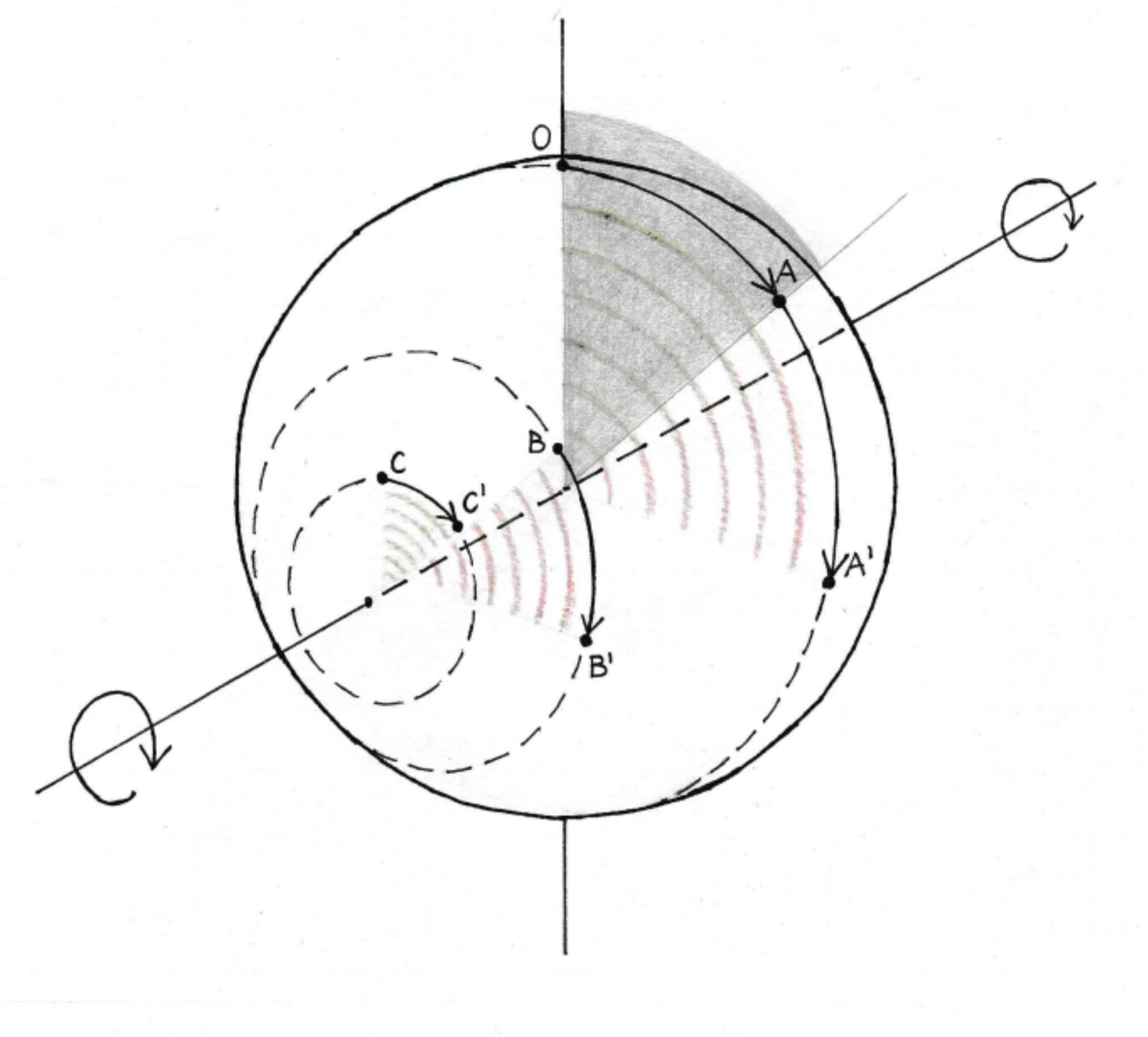}
\end{center}
\caption{\small Action of displacement by a point $A$ on sphere. Displacements by some point are rotations of a sphere in a plane defined by that point, the origin (labeled with 0) and the centre of the sphere, by an amount equal to its polar angle. In particular, $A'=d(A)A=A\oplus A$, $B'=d(A)B=B\oplus A$ and $C'=d(A)C=C\oplus A$. As is clear from the picture, displacements are not homogeneous, and do not commute. To obtain formulas for the hyperboloid from the corresponding formulas for the sphere one replaces  $\beta^2p^2\to-\beta^2p^2$ and Euclidean with Minkowskian scalar product.} \label{fig:M2}
\end{figure}

\begin{equation}
d_S(k)p=p\oplus_S k=\frac{1}{1+\beta^2pk}\left[p\sqrt{1-\beta^2k^2} + k\left( 1+\frac{\beta^2pk}{1+\sqrt{1-\beta^2k^2}} \right)  \right],
\end{equation}
where subscript $S$ stands for Snyder. This expression was first derived in \cite{golf1} using the methods of projective geometry. Its physical meaning is seen as the modification of the energy-momentum conservation law. A particle of momentum $p$ absorbing a particle of momentum $k$ obtains momentum $q=d(k)p=p\oplus k$ (see figure 2). Namely, it is the geometry of energy-momentum background that defines the law of energy-momentum conservation, and vice versa (see the discussion in \cite{relloc}). Only for the flat spaces is this law trivial, while for nontrivial geometries it gets generalized.

Using the fact that for a vector function $f(p)$

\begin{equation}
d(f(k))p=p\oplus f(k),
\end{equation}
and

\begin{equation}
d(k) f(p)=f(d(k)p)=f(p\oplus k),
\end{equation}
it is possible to obtain the addition rule for the orthogonal projection defined by $p=\eta$,

\begin{equation}
d(k)_M p=p\oplus_M k= \left(\sqrt{1+\beta^2k^2}+\frac{\beta^2(pk)}{1+\sqrt{1+\beta^2p^2}}\right)p+k,
\end{equation}
where subscript $M$ stands for Maggiore, who first considered this parametrization \cite{magg}. From it then follows addition rule for a generic projection/parametrization

\begin{multline}\label{gen add rule}
d(k)p=p\oplus k=g\left(\beta^2\left[  h_pp+h_kk\left(\sqrt{1+\beta^2p^2h_p^2}+\frac{\beta^2h_ph_k(pk)}{1+\sqrt{1+\beta^2k^2h_k^2}}   \right)  \right]^2\right) \\
\left[
h_pp+h_kk\left(\sqrt{1+\beta^2p^2h_p^2}+\frac{\beta^2h_ph_k(pk)}{1+\sqrt{1+\beta^2k^2h_k^2}}   \right)\right] ,
\end{multline}
where $h_p=h(\beta^2p^2)$, $h_k=h(\beta^2k^2)$, and $g$ is a function of a complicated argument. 

\begin{figure}
\includegraphics[trim={2.5cm 25cm 0 0},clip]{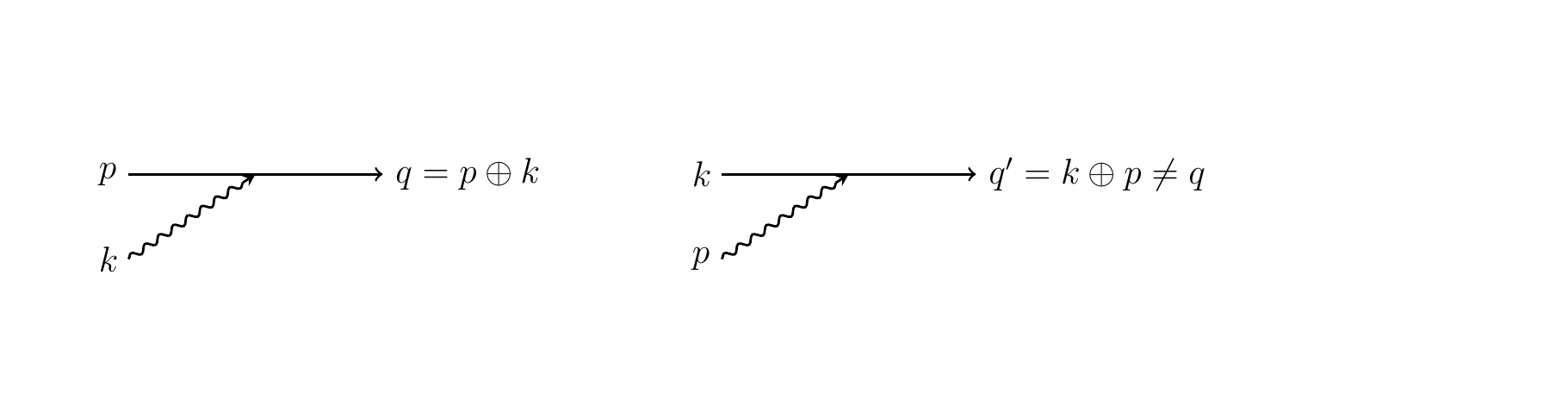}
\caption{\small Elementary manifestation of the noncommutativity of the momenta addition. An electron with a momentum $p$ annihilating a photon of a momentum $k$ emerges with a different energy-momentum than an electron of a momentum $k$ annihilating a photon of a momentum $p$.}\label{fig3}
\end{figure}

Even though one can switch back and forth between different projections/reparametrizations, they describe very different physics. For instance, displacements for the homogeneous projection are an isometry, in the sense of $d(p\oplus k,q\oplus k)=d(p,q)$, for any three points $p$, $q$ and $k$. This is not so for the orthogonal projection. Physically, this leads to very different consequences. In the first case, two electrons of the momenta $p$ and $q$ remain on the same geodesic distance upon absorption of the momenta $k$ from the electromagnetic field, while in the latter, they approach each other or move away in the same situation.

It is only for flat space that the generators of displacements are identical to the points on the cotangent manifold. On curved spaces this is no longer so, due to the non-equivalence of cotangent spaces at different points. This means that the space-time manifold differs from point to point (is a function of $p$), and is not the same as the space of displacement operators $\tilde x$, unless $p=0$. In fact, elements of the space-time manifold can be expressed in terms of displacement operators via

\begin{equation}\label{x}
x(p)=\frac{\partial}{\partial p}=\frac{1}{f_1}\hat x -\frac{\beta^2f_2}{f_1(f_1+f_2\beta^2p^2)}p(p\hat x),
\end{equation}
with $x(0)=\hat x$.
The space-times belonging to two points $p$ and $q$ are related via,

\begin{equation}
x(p)=\tau(p,q)x(q),
\end{equation}
where transport operator is given as

\begin{equation}
\tau(p,q)=\frac{f_{1q}}{f_{1p}}\mathbf{1}+\beta^2\frac{f_{2q}}{f_{1p}f_{1q}^2}q\otimes q-\beta^2\frac{f_{2p}f_{1q}}{f_{1p}(f_{1p}+f_{2p}\beta^2p^2)}p\otimes p-\beta^2\frac{\beta^2(pq)f_{2p}f_{2q}}{f_{1q}^2f_{1p}(f_{1p}+f_{2p}\beta^2p^2)}p\otimes q,
\end{equation}
where $f_{ip}=f_i(\beta^2p^2)$, $f_{iq}=f_i(\beta^2q^2)$, $i=1,2$. This forms the essence of the relative locality principle \cite{relloc}. The expression for the infinitesimal

\begin{equation}
dx(p)=\left(\frac{\partial \tau}{\partial p}dp     \right)x(q)+\left(\frac{\partial \tau}{\partial q}dx     \right)x(q)+\tau dx(q)
\end{equation}
shows that locality is not absolute: a certain space-time event for an observer of momentum $p$ generally does not necessarily map into a single point for an observer of different energy-momentum $q$, that is $dx(p)=0$ does not imply $dx(q)=0$, unless energy-momentum is flat ($\tau=\mathbf 1)$.

The fact that global infinitesimal displacement operators differ from cotangent vectors, and that cotangent vectors at different points of the energy-momentum manifold can not be identified, is the main obstacle to the formulation of space-time dynamics corresponding to curved energy-momentum space. One proposal is to simply identify space-time with infinitesimal displacements operators, as was originally proposed in Snyder's seminal work \cite{Snyder1}. In this case standard dynamics, such as in e.g. Hamiltonian formalism, is to be modified by replacing flat momenta $p$ with the curved ones, and coordinates with generalized infinitesimal displacements, the approach that was followed in e.g \cite{mign2, hydsn}.  Another proposal is to formulate dynamics in terms of local canonical coordinates $x(p)$, "the momenta of the momenta", as in the relative locality framework \cite{relloc}. Yet another possibility is to Fourier transform momentum space equations, with eigenfunctions of the Laplace-Beltrami operator for flat space, $e^{-ipx}$, that is, representation of the group of isometries (motions) on flat space, replaced with representations of the same group on $d$-dimensional hyperboloid.\footnote{For a comprehensive exposition of these representations on $d$-dimensional maximally symmetric spaces see \cite{lima}.} This approach was deployed in e.g. \cite{mk2,mk3}. It leads to a theory defined on the lattice, with differential operators replaced with finite difference operators of a step $1/\beta$. As claimed in \cite{kady5}, such theory corresponds to a covariant formulation of the Wilson's gauge theory on the lattice \cite{wil}. Even though this choice is mathematically consistent with the idea of Fourier transform, physically there is a problem in interpretation of the Fourier transformed space. In \cite{kady1, kady5} discrete eigenvalues of the Laplace-Beltrami operator were interpreted simply as the granular configuration space (space-time), while in \cite{mk4} it was called $\xi$ representation, and another transform of a Fourier type was preformed on the physical quantities (fields) to obtain their space-time dependence. Finally, there is a choice to consider dynamics in the embedding Minkowskian space \cite{kady1,kady2,kady3,kady4}, for which Fourier transform is the standard one, but where each equation of motion is accompanied with an equation that constraints particle momenta to lie on the embedded surface. 

 We shall avoid this conundrum by studying simple physical systems entirely in the energy-momentum representation, and try to extract as much physical content out of it. This approach is justified by a complete equivalence of space-time and energy-momentum representations for the description of physics in the standard, flat case. As one can switch back and forth between the two representations by a simple Fourier transform, neither can be considered more fundamental, i.e. it should be the same whether one uses space-time representation or energy-momentum representation as a starting point for the modification of the laws of physics. Using the latter avoids the above mentioned ambiguities that come with the modification of the dynamics at short length scales. 

\subsection{Generalized Dirac's delta}
The concept of Dirac's delta function is readily generalized to an arbitrary background, with a natural definition

\begin{equation}
\int f(p) \delta(p\ominus k) d\Omega_p=f(k).
\end{equation}
This was considered in \cite{mk1, mk2, mk5, mign2}. From it immediately follows the relation with the standard (flat space) Dirac delta,

\begin{equation}\label{delta}
\delta(p\ominus k)=\frac{1}{\sqrt{\det g}}\delta(p-k).
\end{equation}
This can be demonstrated by elementary means, via

\begin{equation}
\delta(p\ominus k)=\frac{1}{\left|\det \partial_p(p\ominus k)|_{p=k}\right|}\delta(p-k),
\end{equation}
using the fact that for multidimensional arguments the delta function has the property

\begin{equation}
\delta(f(x))=\sum_{i=1}^n \frac{1}{|\det \partial f |_{x=x_{0i}}   |}\delta(x-x_{0i}),
\end{equation}
where $x_{0i}$ are the $n$ zeroes of the function $f$, as well as that for isotropic projections from a maximally symmetric space the antipode, $\ominus p$, defined by

\begin{equation}
(k\oplus p)\oplus(\ominus p)=k
\end{equation}
for every $k$, is unique and trivial, $\ominus p=-p$ (as follows from the geometry, see fig. 1).

%
Therefore,

\begin{equation}
\partial_p (p\ominus k)|_{p=k}=h_k\mathbf 1_-+\frac{\beta^2h_k^3}{1+h_k}k\otimes k,
\end{equation}
whose determinant squared equals the determinant of the corresponding metric,

\begin{equation}
g(k)=\frac{(1-\beta^2k^2)\mathbf 1_-+\beta^2k\otimes k}{(1-\beta^2k^2)^2}.
\end{equation}
 The proof that this relation holds in general case is left as an exercise in cumbersome algebra.

\section{Physical examples}\footnote{In this section the gauge is fixed so that $(pA(p))=0$.}
\subsection{Plane-waves}
The solutions to the Maxwell equations outside sources are plane waves. These are given as $e^{ikx}$ in the coordinate, and as $\delta(p-k)$ in the momentum represantation. Following the general procedure, it is immediately noticable that the solutions in the momentum space of non-zero curvature are generalized Dirac's delta's, i.e.

\begin{equation}
A(p)=(2\pi)^{d/2}\delta(p\ominus k).
\end{equation}
This follows from the defining relation of the delta function and equation (\ref{momentum maxwell standard}) upon integration. Thus in our minimal extension prescription, the plane waves are the same as in the flat case, only ordinary delta function is replaced by a generalized form in (\ref{delta}). The result depends on the projection only through the delta function.

\subsection{Point charge source}
Consider a field of a static point charge situated at the origin, $j(x)=j(\mathbf{r})=(e\delta(\mathbf{r}),0,0, 0)$, where $x=(t,\mathbf{r})$ and for definiteness we take 3+1 space-time dimensions . In the momentum representation, this becomes

\begin{equation}
j(\mathbf p)=(e,0,\dots 0).
\end{equation}
Following our principle of minimal extension, as defined in section 2, we take that on passing to energy-momentum background of constant curvature, the current vector is a constant vector, on curved background. Maxwell equation has only zero-th component, from which is read out scalar potential

\begin{equation}
\phi(\mathbf p)=\frac{e}{\mathbf p^2},
\end{equation} 
which in turn gives the electric field strength

\begin{equation}
\mathbf{E(p)}=\mathbf{p}\phi(\mathbf{p})=\frac{e}{|\mathbf p|}\mathbf{\hat p},
\end{equation}
where $\mathbf{\hat p}$ is a unit vector in the direction of $\mathbf{p}$. From here, the total energy of the static point charge field configuration is

\begin{equation}
\mathcal E=\frac{1}{4\pi}\int \mathbf{E}^2(\mathbf p)d\Omega_{\mathbf{p}}=\alpha\int \frac{d\Omega_{\mathbf{p}}}{|\mathbf{p}|^2},
\end{equation}
where $\alpha$ is a fine-structure constant.\footnote{In the standard, flat case, the volume element is trivial $d\Omega_p=d^3p$, and the energy integral is linearly divergent in the UV. Compared to the linear divergence as $r\to 0$, this point illustrates nicely the relation between short-scale configuration and large scale momentum representations.} The three-momentum part of the surface is a hypersphere $S^3$, so for the homogeneous coordinates

\begin{equation}
\int d\Omega_\mathbf{p}=\int_0^\infty\frac{4\pi p^2dp}{(1+\beta^2p^2)^2},
\end{equation}
giving finite total self-energy

\begin{equation}
\mathcal E=\pi^2\alpha\beta^{-1}.
\end{equation}

We note that this result is not independent of the projection. For instance, taking the case of orthogonal projection

\begin{equation}
\int d\Omega_\mathbf{p}=\int_0^\infty\frac{4\pi p^2dp}{(1+\beta^2p^2)^{1/2}},
\end{equation}
gives an infinite self-energy of a point charge. This can be related to the fact that the total volume of space for orthogonal projection is infinite, while that of a homogeneous projection is finite.\footnote{The importance of the finitness of the total volume of space was recognized as crucial for the finitenes of the quantum field theory already in \cite{golf1}.} Thus, it is homogeneous projection that preserves finitness of the total volume (surface) of the hypershere (in addition to isometries), while orthogonal does not. This clearly demonstrates that the theory is not reparametrization invariant, as claimed in some papers \cite{relloc}. 

As a final exercise, we may set a bound on the value of curvature parameter $\beta$. Taking the model of electron as homogeneously charged ball of radius $R$, the total value of it's electromagnetic energy is

\begin{equation}
\mathcal E=\frac{3}{5}\frac{\alpha}{R},
\end{equation}
from which is inferred $\beta\sim R$. The bound on an electron radius from \cite{deh}, sets the bound on the curvature parameter

\begin{equation}
\beta < 10^{-20}\ \text{cm},
\end{equation}
while a more stringent bound can be obtained from an indirect estimate of electron's radius, such as through the electron's dipole moment strength, via $d\sim eR$, where the most recent measurements \cite{edm} constraint

\begin{equation}
\beta < 10^{-29}\ \text{cm},
\end{equation}
which is near (within couple orders of magnitude) Planck's length, $l_p\sim 10^{-33}$ cm, a scale that is usually inferred as the threshold of new physics.

\section{Conclusion and outlook}
In this paper was considered the simplest possibility of generalizing Maxwell's theory to incorporate curvature of the energy-momentum space. Such "minimal extension" conserves the symmetries of the theory, the Lorentz and gauge symmetry. In the process, we demonstrated non-equivalence of different parametrizations of the momentum manifold. For an isometry preserving projection, such as homogeneous projection defined in (\ref{sphere hom proj}), the point charge self energy is shown to be finite. Compared to other extensions that lead to the same result, such as Born-Infeld model \cite{BI}, the model considered here is simpler in that it does not require the addition of extra terms in the action, therefore providing a better physical explanation for it's input. 

The necessity of a modification of a fundamental space-time structure at the shortest distances to incorporate quantum gravity is a well established fact, as is properly recognized corresponding generalization of the energy-momentum geometry, see e.g. \cite{relloc}. Still, it is our impression that the study of geometry of the energy-momentum space comes only as a step towards the description in space-time, and that the abundance of physical content that it holds within itself has not been enough exploited in the literature. In this sense, the obtained result of the finiteness of the point charge self-energy, serves as the principal justification for our approach, and a strong motivation for further research in this direction.

While the energy-momentum representation is not well suited for the study of classical
dynamics, which is given in terms of particle trajectories and field configurations in space and time, it is still possible to extract certain global physical information about the system from it, as has been shown in this paper for the case of a point charge. The next step would be to apply the formalism in the case of a quantum and quantum field theory. Being involved with calculation of the spectra of operators, decay rates and cross sections, quantum theory is manifestly independent of the representation (Born reciprocity) and thus much more suited for the application of the minimal extension principle. This was considered a long time ago in \cite{golf1,Golf}, but it also has certain justification even from flat space QED. Namely, there is a much better interpretation in terms of photons wave (probability) function of $A(p)$ then $A(x)$, see e.g. \cite{qed}. Finally, we emphasize that the study of a particular case of electron's self-energy in the classical context has consequences for the quantum case as well, due to the correspondence principle between the two, as noted in \cite{vf}.


\section*{Acknowledgment}
We are grateful to S. Mignemi for careful reading of several versions of rough draft, to S. Fredenhagen for an illuminating discussion, and to  T. Kalac for drawing the figure.

\end{document}